\documentstyle[preprint,prd,aps,epsfig]{revtex}
\begin{document}

\title{ Magnetic Deformation of Atoms in the Crust of Magnetars}

\author{Nandini Nag $^{a)}$\thanks{E-mail:nandini@klyuniv.ernet.in}
and Somenath Chakrabarty$^{a),b)}$\thanks{E-mail:somenath@klyuniv.ernet.in}\\
a)\thanks{Permanent address}
 Department of Physics, University of Kalyani, West Bengal 741 235,
India\\
b) IUCAA, Post bag 4, Ganeshkhind, Pune 411 007, India}

%\date{\today}
\maketitle

\begin{abstract}
Some special features of magnetically deformed  multi-electron atoms 
in the crustal region of strongly magnetized neutron stars (magnetars)
are studied with  the relativistic version of semi-classical
Thomas-Fermi model. An analytical solution of Poisson's equation with
cylindrical symmetry is obtained. The variation of atomic dimension with 
magnetic field strength is also studied. An expression for atomic quadrupole
moment, appears because of magnetic deformation is derived. The effect
of atomic
quadrupole-quadrupole interaction in the equation of state  of crustal
matter of magnetar is briefly discussed. 
\end{abstract}
\newpage

	From  the observational evidence of some compact stellar objects
with extremely high surface magnetic field \cite{R1}, the study of the effect
	of ultra strong magnetic field on the geometrical structure
	(electronic orbits) of atoms in the crustal region have got a new 
	dimension. The
	possibility of such ultra-high  magnetic field in these stellar
	objects has given re-birth to this "the old subject" of quantum
	mechanics. These strongly magnetized objects are known as
magnetars \cite{R2}. The measured value of magnetic field strength at the
	magnetar surface can be as high as $\geq 10^{15}G$. These objects 
	are also supposed to be
	the sources of soft gamma repeaters and anomalous X-ray pulsars.
	Soft gamma repeaters are a class of compact stellar objects
	which periodically emit bursts of gamma rays, in addition to
	persistent X-rays. These two classes of objects appear to be
	very similar (aside from the presence of bursts in SGR's). The
	measured period lie in the range $5-12$ secs. The dipole fields
	at the surface inferred from the spin down rate are
	$10^{14-15}G$, which are much larger than the surface field
	strength of
	 known strongly magnetized conventional neutron stars ($\simeq
	10^{12}G)$. It has long been suggested that strong magnetic
	energy may be the ultimate source of both the bursts and the
persistent radiation \cite{R2}. From the observational data it is believed
	that such objects (magnetars) are strongly magnetized young
	neutron stars. If the strength of surface magnetic field is
	$\geq 10^{15}G$, then it would be interesting to investigate the effect
	of such strong magnetic field on the properties of crustal matter. 
	Some of the properties of such low density matter in presence of
strong magnetic field have already been studied \cite{R3}. The other interesting
effect of ultra-strong magnetic field on the crustal matter, which has
not been explored in greater detail, which we are going to discuss in this
article  is the possibility of magnetic deformation of atoms
	at the lattice points in the crustal region. Although the
physical problem related to the atoms placed in an external magnetic field 
is one of
	the oldest subject of quantum mechanics, this particular exotic
	phenomena -- the magnetic deformation of atoms, heavier than
	hydrogen, have not been studied. The effect of strong magnetic
	field on the stability of hydrogen atoms and on the multi-proton
bound states have been investigated in detail \cite{R4,R5,R6,R7}. In a recent 
work, the magnetic deformation of hydrogen atom and the appearance of
its quadrupole moment have been studied \cite{R8}.  In the present report, 
we shall consider the effect of strong magnetic field on the geometrical
structure of multi-electron atoms (much heavier than hydrogen atom).
We do believe that such physical situation can not be ruled out 
at the crustal region of magnetars. We shall also investigate the
        effect of strong magnetic field on the  quadrupole moment of 
deformed atoms at the crustal lattice points.
It is believed that the shape of an atom is almost spherical at $B
	\leq B_0$ and a cylindrical at $B\geq B_0$, where
	$B_0=m_e^2e^3=2.35\times 10^9 G$, $m_e$ and $e$ are respectively
	the mass and electric charge of the electron (in our calculation
	we have assumed $\hbar =c=1$). The critical parameter
	$\gamma=B/B_0$ has significance in the non relativistic theory. In our
	case $B\geq 4.4\times 10^{13}G$ which is the relativistic region
	and we consider the parameter $\zeta=B/B_c$, where
	$B_c=4.4\times 10^{13}G$, the typical field strength at
	which the Landau levels of electrons are populated.
Since  fully relativistic calculation of a many body quantum mechanical 
system of finite
size in presence of ultra-strong magnetic field is a very difficult
task, even numerically, we are considering for the
	multi-electron atoms in presence of strong magnetic field
$B \geq B_c$, a simple semi-classical approach- the relativistic version
of Thomas-Fermi model in presence of strong magnetic fields.

We shall now consider the basic formalism of relativistic Thomas-Fermi
model in presence of strong magnetic field.
The electrostatic potential $\phi$ satisfies the equation \cite{R3},
	\begin {equation}
	   \nabla^2\phi=4\pi n_ee-4\pi Ze\delta(r-r_n)
	   \end{equation}
	        Where $n_e$ is the electron density, $Z$ is the atomic
	number and $r_n$ is nuclear radius. Since $r_n\sim$ fm $\ll$ atomic
radius $\sim A^0$, the nuclear contribution on the right hand side can
therefore be neglected for $r > r_n$. Now assuming that only  the
zeroth Landau level for the electrons is populated, the chemical potential
	of electron is given by 
	\begin{equation}
	\mu=({p_F}^2+{m_e}^2)^{1/2}-e\phi(r)={\rm{constant}}
	\end{equation}
	This gives,
	\begin{equation}
	p_F=[(\mu+e\phi)^2-{m_e}^2]^{1/2}
	\end{equation}
	the electron Fermi momentum. 
	Now in presence of a strong quantizing magnetic field of strength
	B, the electron number density is given by,
	\begin{equation}
	n_e=\frac{eBp_F}{\pi^2}=\frac{eB}{\pi^2}[(\mu+e\phi)^2
	-{m_e}^2]^{1/2}
         \end{equation}
	In the  high density region, we can neglect $m_e$, then 
	\begin{equation}
	n_e \simeq \frac{eB}{\pi^2}(\mu+e\phi)
	\end{equation}
	and eqn.(1) becomes
	\begin{equation}
	\nabla^2\phi=\frac{4 e^2 B}{\pi}(\mu+e\phi)
	\end{equation}
	Substituting $\mu+e\phi=\psi$, we have in the cylindrical
	coordinate $(r,\theta,\phi)$
	\begin{equation}
	\psi_{rr}+\frac{1}{r} \psi_r+\frac{1}{r^2}
     	 \psi_{\theta \theta}+\psi_{zz}=\alpha\psi
	\end{equation}
	Where, $\psi_x=\partial\psi/\partial x$
	and
	\begin{equation}
	\alpha=\frac{4Be}{\pi}
	\end{equation}
	Assuming cylindrical symmetry, $\psi$ becomes independent of
	$\theta$. Then with the separable form of solution
	\begin{equation}
	\psi(r,z)=R(r)Z(z)
	\end{equation}
	we have
	\begin{equation}
	\frac{d^2 Z}{dz^2} -s^2Z=0
	\end{equation}
	and,
	\begin{equation}
	\frac{d^2 R}{dr^2}+\frac{1}{r}\frac{dR}{dr}-p^2R=0
	\end{equation}
	Where,  $p^2=s^2=\alpha/2$.
	The solution of the first equation is
	\begin{equation}
	Z(z)={\rm{constant~}}\times \exp(-sz)
	\end{equation}
	       Replacing $r$ by $\bar{r}=pr$, we have from the second
	       equation
	\begin{equation}
	\frac{d^2 R}{d\bar{r}^2}+\frac{1}{\bar r}\frac{dR}{d\bar{r}} -R=0
	\end{equation}
	The solution of this equation is given by,
	\begin{equation}
	R(r)={\rm{constant~}} \times K_0(\bar{r})
	\end{equation}
	where $K_0(\bar{r})$ is the second kind modified Bessel function
	of order zero. Therefore,
	\begin{equation}
	\psi (r,z) =C \times \exp(-sz)K_0(\bar{r})
	 =C \times \exp(-\bar{z}) K_0(\bar{r})
	 \end{equation}
	 To obtain the normalization constant $C$, we use the relation,
	 \begin{equation}
	 \lim_{l \rightarrow r_n}\psi=\lim_{l \rightarrow
	 r_n}e\phi l=Ze^2
	 \end{equation}
	 Where $l=(r^2+z^2)^{1/2}$, some arbitrary distance from the
 centre of the nucleus to the surface of the cylinder. Since 
 $r_n \ll r_{{\rm{atom}}}$, we can approximate  the normalization
 constant by
	 \begin{equation}
	 C \simeq \frac{Ze^2}{r_nK_0(\bar r_n)}
	 \exp(\bar r_n)
	 \end{equation}
which is $\approx 10^{11}$ for iron atoms and is insensitive with the
variation of magnetic field strength.
Since $r=\bar{r}/p$ and $z=\bar{z}/s$, we can conclude that both the 
longitudinal and lateral dimensions of the deformed atom decrease
	 with the increase in magnetic field strength. To obtain 
the surface of a deformed atom we use the relation,
	 \begin{equation}
	 \frac{\partial \psi}{\partial l}=0
	 \end{equation}
	 which means the electric field vanishes at the surface. Hence we
	 get,
	 \begin{equation}
	 \exp(-\bar{z})[\bar{r}K_0(\bar{r})+\bar{z}K_1(\bar{r})]=0
	 \end{equation}
Solving this equation we obtain $r$ as a function of $z$. We
	 have noticed that the atoms become ellipsoidal with cylindrical
	 symmetry in presence of strong magnetic field. The variation of
	 $r$ with $z$ is given by the equation
	 \begin{equation}
	 \frac{z^2}{a^2} +\frac{r^2}{b^2}=1
	 \end{equation}
The length of both the major and minor axes decrease with the increase
of field strength. The variations are given by the following power laws.
	 \begin{equation}
	 a=0.53 \zeta^{-0.5}
	 \end{equation}
	 \begin{equation}
	 b=0.436 \zeta^{-0.501}
	 \end{equation}
	 From these power law relations it is very easy to see that the
	 decrease of $b$ with $B$ is slower than that of $a$.
	 This is also obvious from the asymptotic nature of $K_0(\bar
	 r)$ which is given by \cite{RR8}
	 \begin{equation}
	 K_0(\bar{r})
	 \sim \frac{\exp(-\bar{r})}{(\bar{r})^{1/2}}
	 \end{equation}
for large $\bar{r}$. Which means the deformed atom maintains its
ellipsoidal shape even for very high value of magnetic field strength.
The quadrupole moment for the deformed atom with ellipsoidal shape of
	 the type  given by eqn.(20) may be written as \cite{R8,R9,R10}
	 \begin{equation}
	 Q_{ij}=\sum  \rho(2x_ix_j-l^2\delta_{ij})
	 \end{equation}
where the sum goes over all the charges. Now for the charge distribution 
with cylindrical symmetry 
	 \begin{equation}
	 Q_{zr}=Q_{rz}=0
	 \end{equation}
	 and
	 \begin{equation}
	 Q_{zz}=-Q_{rr}=\pi \int \rho(z,r) (z^2-r^2)r^2 dz
	 \end{equation}
	 where,    
	 \begin{equation}
	 \rho(r,z) \simeq \frac{eB}{\pi^2} \psi(r,z)
	 \end{equation}
	 Expressing $r$ in terms of $z$ using eqn.(20), we have obtained 
	 $Q_{zz}$ as a
	 function of $\zeta$. The variation of $Q_{zz}$ with $\zeta$ is given
	 by the power law,
	 \begin{equation}
	 Q_{zz}=C \times 0.029 \zeta^{-1.5}
	 \end{equation}
for $\gamma >> 1$, $C$ is the same normalization constant as mentioned
before. However, for a very broad range of $\zeta$, the dependence is
completely different, particularly for low $\zeta$, In fig.(1) we have
shown the change of $Q_{zz}$ as a function of $\zeta$ for a very wide
range. The variation is almost identical with the recent result on
deformed hydrogen atom reported by Potekhin \cite{R8}.

We shall now consider a deformed atom in presence of external electric
field produced by the quadrupole moment of other deformed atoms. Let us
assume that there are two such systems, each having total charge zero
and vanishing dipole moment.
Then we have the total potential energy of the system
\begin{equation}
U=\sum_i e_i \phi_i(\vec{r_i})
\end{equation}
where $\vec{r_i}$ is the radius vector of the charge  $e_i$ with
its origin anywhere within the system. Since both the total charge and
dipole moment of the system are zero, we have the total potential energy
due to quadrupole field
\begin{equation}
U^{(2)}=\frac{Q_{\alpha\beta}}{4}
\frac{\partial^2\phi^{(2)}}{\partial X_{\alpha}\partial X_{\beta}}
\end{equation}
where $\phi^{(2)}$ is quadrupole field and $Q_{\alpha\beta}$ is the
quadrupole moment. The quadrupole field is given by,
\begin{equation}
\phi^{(2)}=\frac{3}{4} Q_{\sigma\delta}
\frac{X_{\sigma}X_{\delta}}{R_0^5}
\end{equation}
which is the quadrupole field produced by the quadrupole moment
$Q_{\sigma \delta}$ of a deformed atom at the point with radius vector
$\vec{R_0}$.
Combining these two equations we have after some straight forward
algebra,
\begin{equation}
U^{(2)}=\frac{Q_{zz}^2}{R_0^5} \left [\frac{105}{16R_0^4}(z^2-r^2)^2-3
\right ]
\end{equation}
This is the potential energy of the deformed atom in presence of the
quadrupole potential $\phi^{(2)}$ produced by another deformed atom,
i.e., it is the extra two body potential arises because of
quadrupole-quadrupole interaction.
% FIGURE
\begin{figure} 
\psfig{figure=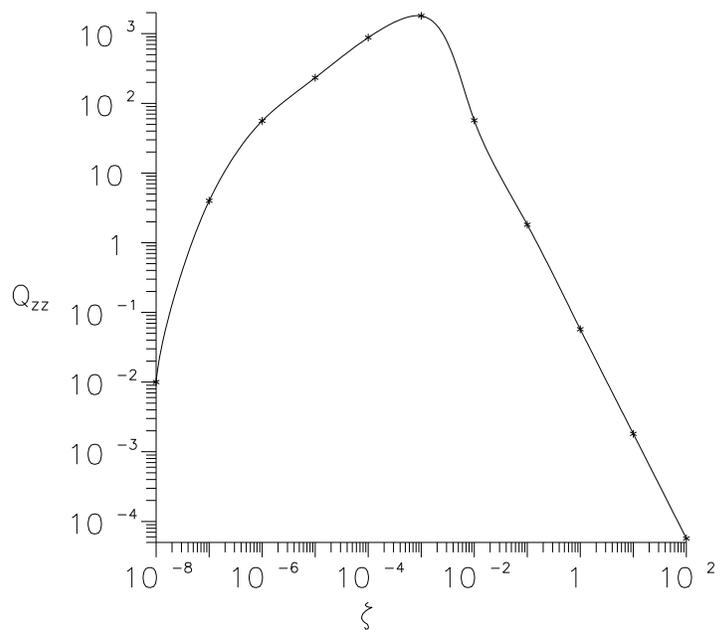,height=0.5\linewidth}
\caption{Variation of quadrupole moment (in $\AA^2$) with the
parameter $\zeta=B/B_c$, the scaling parameter is the normalization
constant $C\approx 10^{11}$.}
\end{figure}

%REFERENCES

	 \end{document}